# Observation of Berry Curvature-Enhanced Anomalous Photo-Nernst Effect in Magnetic Weyl Semimetal

Zipu Fan,[1,*] Jinying Yang,[2,3,*] Yuchun Chen,[1] Ning Zhao,[4] Xiao Zhuo,[1] Chang Xu,[1] Dehong Yang,[1] Jun Zhou,[4] Jinluo Cheng,[5] Enke Liu,[2] and Dong Sun[1,6,7,†]

[1]International Center for Quantum Materials, School of Physics, Peking University, Beijing 100871, China
[2]Beijing National Laboratory for Condensed Matter Physics, Institute of Physics, Chinese Academy of Sciences, Beijing 100190, China
[3]School of Physical Sciences, University of Chinese Academy of Sciences, Beijing 100190, China
[4]School of Physics and Technology, Nanjing Normal University, Nanjing 210023, China
[5]GPL Photonics Laboratory, Key Laboratory of Luminescence Science and Technology, Chinese Academy of Sciences & State Key Laboratory of Luminescence Science and Applications, Changchun Institute of Optics, Fine Mechanics and Physics, Chinese Academy of Sciences, Changchun 130033, China
[6]Collaborative Innovation Center of Quantum Matter, Beijing 100871, China
[7]Frontiers Science Center for Nano-optoelectronics, School of Physics, Peking University, Beijing 100871, China
[*]These authors contributed equally: Zipu Fan, Jinying Yang
[†]Email: sundong@pku.edu.cn

## Abstract

The anomalous Nernst effect is the thermoelectric counterpart of the anomalous Hall effect, which can emerge in magnetic materials or topological materials without magnetic field. Such effect is critical to both fundamental topological physics and various application fields, including energy harvesting, spintronics and optoelectronics. In this work, we observe the anomalous photo-Nernst effect, which use light excitation to generate temperature gradients for the thermoelectric response. Our experiments reveal a pronounced edge photocurrent response in magnetic Weyl semimetal $Co_3Sn_2S_2$ under zero magnetic field, originating from the anomalous photo-Nernst effect. The pronounced photo-Nernst current benefits from the exceptional properties of $Co_3Sn_2S_2$, including the large thermoelectric coefficient, topologically enhanced anomalous response of Weyl bands and the Shockley–Ramo nature of long-range photocurrent generation. Furthermore, by comparing the nominal anomalous Nernst coefficient under different wavelength excitations, we observe a clear enhancement in the mid-infrared region, originating from the topological contribution from the large Berry curvature of Weyl bands. Our results reveal the interplay among light, magnetism, and topological order in magnetic Weyl semimetals, which not only offers insights for fundamental physics but also advances potential applications in quantum devices.

## I. INTRODUCTION

The magnetic order of magnetic materials or the quantum geometry of the electronic wavefunction could both result in an anomalous correction to the band velocity of Bloch electrons [1]. Such correction may lead to transverse movement of carriers, where an external magnetic field is otherwise required to detour the carrier movement from the longitudinal motion [2,3]. These effects are usually termed "anomalous". Anomalous transport effects, such as the anomalous Hall effect and the anomalous Nernst effect, are important for the study of fundamental quantum physics and the development of next-generation quantum devices [4]. On one hand, anomalous transport effects have been shown to be closely related to Berry curvature, making it a powerful tool for characterizing the quantum geometry of the electronic wavefunction of topological materials [4,5]. On the other hand, the ability for transverse charge transport in the absence of an external magnetic field also holds promise for innovative electronic and spintronic devices [6,7]. Among these, the anomalous Nernst effect, a transverse thermoelectric phenomenon, has attracted significant attention for its potential applications in thermoelectric devices and thus energy harvesting [8]. Compared with transport measurements of the thermoelectric effect, which typically require a heat source to create a temperature gradient, the photo thermoelectric effect is a counterpart of the thermoelectric effect that uses light excitation as an alternative approach to generate local thermal gradients [9]. Similarly, the counterpart to the Nernst effect with photoexcitation is termed the photo-Nernst effect [10-13].

Experimentally, observing the photo-Nernst effect is not a trivial task. Since the photothermal effect generates only a local temperature gradient, leading to a local photo-Nernst current, it is essential to consider whether this local effect can produce a global photocurrent that is detectable by electric measurement. This challenge is particularly significant in semiconductors, where locally generated carriers typically scatter before they can diffuse to the electrodes if the illumination occurs away from the electrode area [14]. For this reason, very high mobility sample is required to observe a global effect in semiconducting materials [13]. In conducting materials, such requirement is loosened with the global response induced by the Shockley–Ramo theorem [15]. However, to observe the photo-Nernst effect in simple metals is also difficult, as the Nernst response is typically small because of Sondheimer cancellation [16]. Considering the above factors, semimetallic materials are more favorable than metals and semiconductors. On one hand, as semimetals are conducting, the local photo-Nernst effect can generate a global current according to the Shockley–Ramo theorem [10]. On the other hand, many semimetals also benefit from enhanced Nernst coefficients owing to their unique band structures [17-19]. In addition, magnetic Weyl semimetals (WSMs), which have demonstrated a significant anomalous Nernst effect [20-24] due to both their magnetic order and topological properties, also hold promise for observing the anomalous photo-Nernst effect [25].

$Co_3Sn_2S_2$, a well-established ferromagnetic (FM) WSM [26,27], combines the ferromagnetism of traditional FM materials with the exotic topological properties of WSMs. Experimentally, $Co_3Sn_2S_2$ has demonstrated a large anomalous Hall effect and a significant zero-field anomalous Nernst effect [21,27], making it a promising candidate for observing the anomalous photo-Nernst effect. In this work, the anomalous photo-Nernst response is revealed by scanning photocurrent microscopy (SPCM) measurements in $Co_3Sn_2S_2$. In the absence of an external magnetic field, we observed a clear photocurrent at the edge of the sample in the FM phase. When the magnetization direction is reversed, the direction of the edge photocurrent also reversed accordingly, indicating its anomalous photo-Nernst origin, which is further confirmed by temperature-dependent measurements. The agreement between the simulation and the experimental results further confirms the Shockley–Ramo nature of long-range photocurrent generation in our experiments. Furthermore, by comparing the nominal anomalous Nernst coefficient under different wavelength excitations, we observe a significant enhancement in the mid-infrared (MIR) region, revealing the topological contribution from the large Berry curvature of the Weyl bands to the anomalous photo-Nernst response. The topologically enhanced anomalous photo-Nernst response revealed in our work not only helps us probe and understand the topological properties of magnetic WSMs but also opens new possibilities for various applications, such as optoelectronics, spintronics, and thermoelectric energy harvesting.

## II. RESULTS

### A. Edge photocurrent response from the anomalous photo-Nernst effect

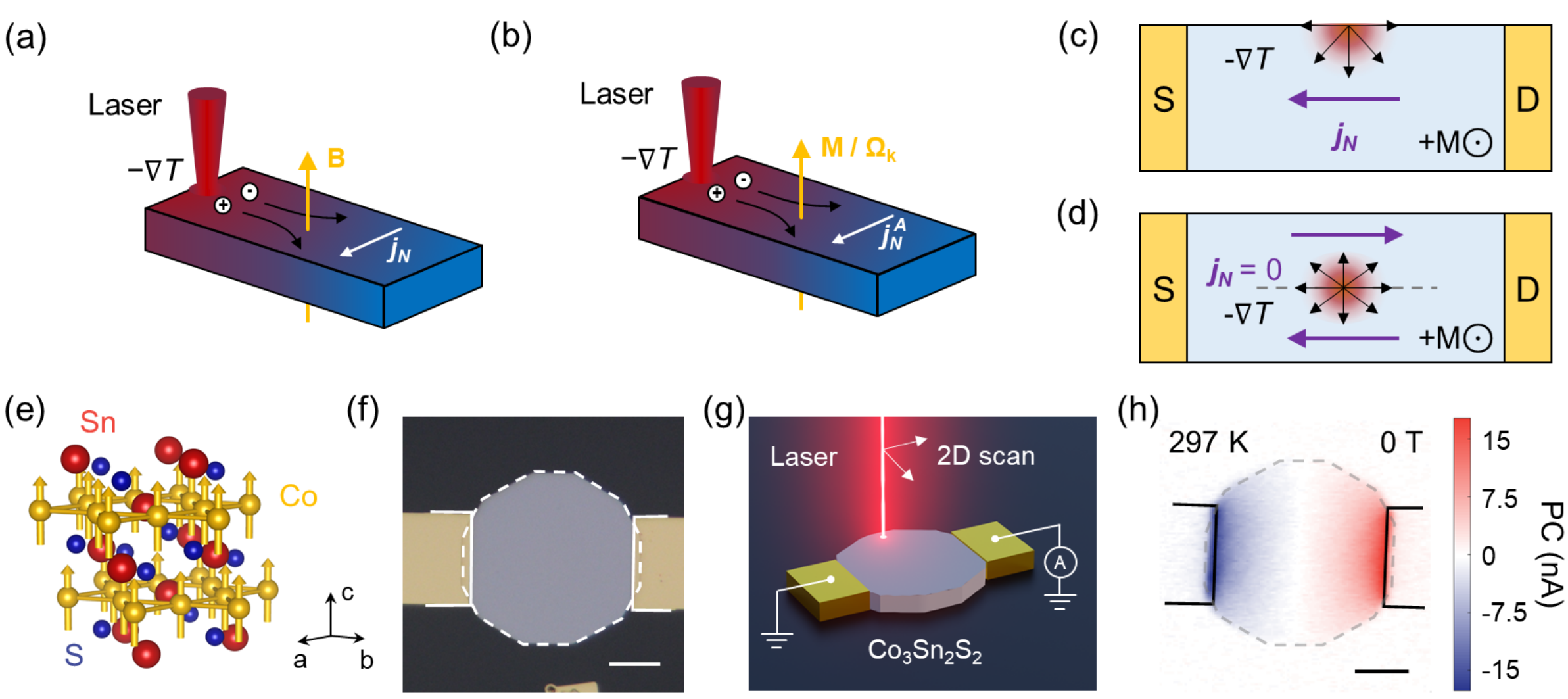

FIG. 1. Schematic for the measurement of anomalous photo-Nernst current. (a), (b) Schematic diagram of the generation of the photo-Nernst current (a) and anomalous photo-Nernst current (b), where the temperature gradient is induced by the laser spot. (c), (d) Schematic diagram of the local generation of the anomalous photo-Nernst current when illuminating the edge (c) and the middle (d) of the sample. The temperature gradient induced by the laser spot is indicated as the red gradient spot. S and D denote the source and drain electrodes, respectively. (e) Crystal structure of $Co_3Sn_2S_2$, showing the stacked ...–Sn-[S-(Co3-Sn)-S]–... layers. The arrows denote the out-of-plane magnetization of $Co_3Sn_2S_2$. (f) Optical microscopy image of the $Co_3Sn_2S_2$ device. (g) Experimental configuration of scanning photocurrent microscopy (SPCM). (h) SPCM image taken at 297 K under 800-nm excitation, with an excitation power of 360 μW.

The schematic illustrations of the photo-Nernst effect and the anomalous photo-Nernst are shown in Figs. 1(a) and 1(b), respectively, where the temperature gradient is generated by light illumination. The easiest scheme for generating a temperature gradient with light excitation is to illuminate the free edge of the sample, as illustrated in Fig. 1(c). In such a scheme, owing to the restriction of the edge, the carriers diffuse away from the edge following the temperature gradient generated by the light beam, and under the influence of the Lorentz force or the Berry curvature [5], a net photo-Nernst current or anomalous photo-Nernst current (APNC) is generated along the free edge. Otherwise, if the light excitation is in the middle area of the sample, as the restriction from the edge is removed, the carriers can diffuse in any direction and the generated photocurrents cancel each other, which is the case illustrated in Fig. 1(d).

$Co_3Sn_2S_2$ is a ferromagnetic (FM) material with a Curie temperature of $T_C \sim 177$ K [28]. When the temperature is below the $T_C$, $Co_3Sn_2S_2$ will develop a ferromagnetic order, which breaks the time reversal symmetry, thus converting $Co_3Sn_2S_2$ into a topological WSM [26,29]. As shown in Fig. 1(e), in the FM phase, $Co_3Sn_2S_2$ has long-range quasi 2D magnetism with out-of-plane magnetization along the *c*-axis [30]. The Weyl node is only 60 meV above the Fermi level, making it accessible through transport experiments. Previously, it has been demonstrated experimentally that a large zero-field ANE can be observed in $Co_3Sn_2S_2$ [21,24,31], and theoretical works have suggested that the large zero-field ANE possibly originates from the topological band structure instead of just ferromagnetic order [19,32]. To study the APNC, we then fabricated an in-plane two-terminal device (Fig. 1(f)) to collect the in-plane photocurrent. The distribution of the photocurrent is studied by performing SPCM measurements, and a schematic of the setup is shown in Fig. 1(g). For comparison, we first performed the SPCM measurement at room temperature under zero magnetic field, where the sample is in the paramagnetic (PM) phase with zero magnetization. As shown in Fig. 1(h), the photocurrent is generated mostly near the contacts, with no significant response at the free edge of the sample.

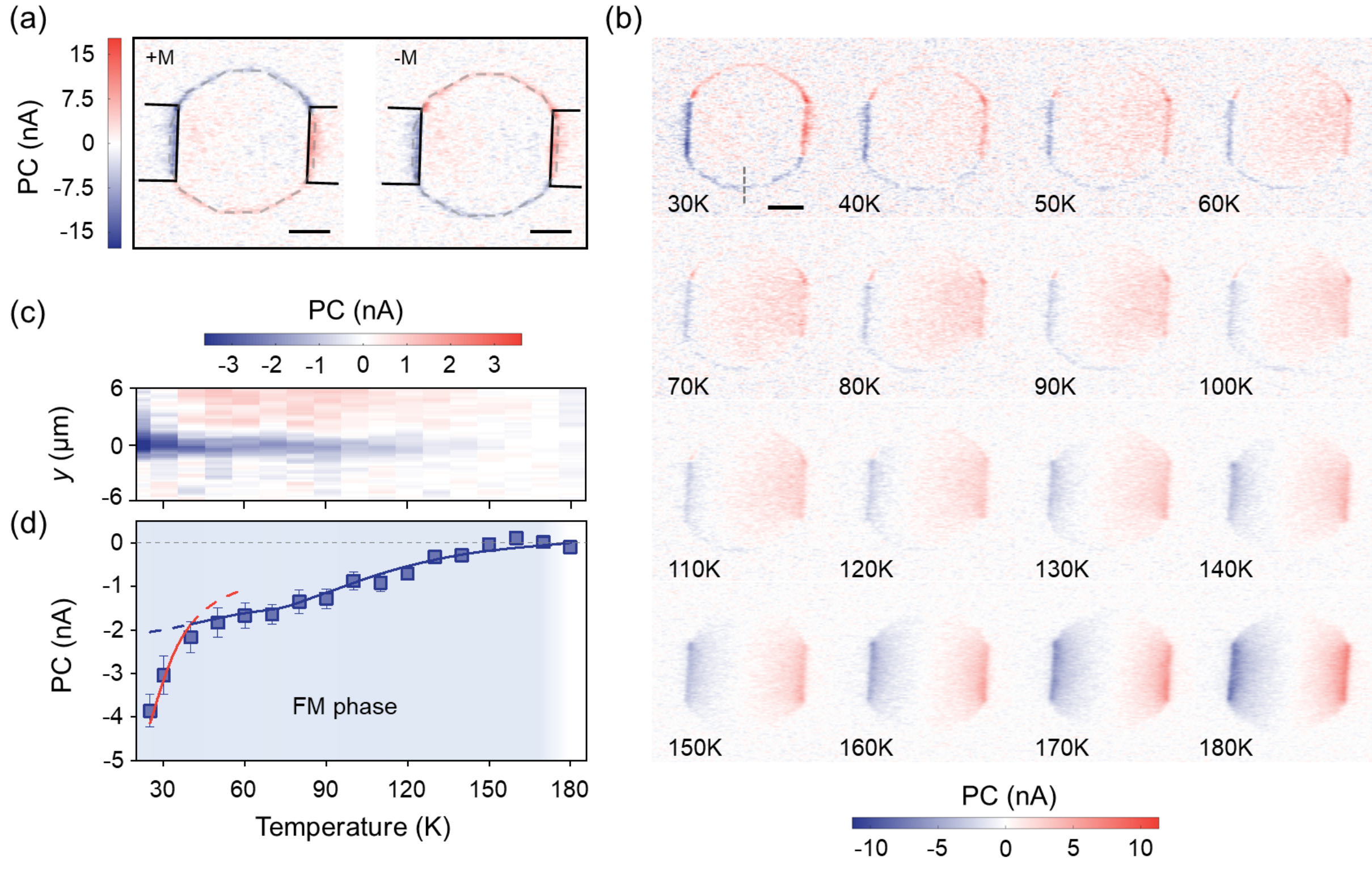

FIG. 2. Anomalous photo-Nernst current in $Co_3Sn_2S_2$. (a) SPCM images taken at 25 K under +*M* magnetization (left panel) and -*M* magnetization (right panel). (b) SPCM images taken at different temperatures under +*M* magnetization. The dashed line in the SPCM at 30 K marks the linecut where the detailed scanning is performed. (c) Detailed scanning results along the dashed line labeled in (b) at different temperatures. (d) The extracted anomalous photo-Nernst current at different temperatures. The error bars represent the standard deviation of 10 scans. The red and blue lines represent the fitting results obtained by adopting different temperature dependences of the thermal contact resistance (red for $T^{-3}$ and blue for $T^{-1}$). For the measurements, the excitation power is fixed at 360 μW. All scale bars are 20 μm.

In contrast, after the sample is cooled below the $T_C$ in a small magnetic field of ~0.1 T (field cooling, FC), when the sample is in the ferromagnetic (FM) phase with out-of-plane magnetization (denoted as ±*M*) following the cooling field (see Supplemental Material Section I for details [33]), a clear photocurrent response at the free edge of the sample is observed, and the sign of the photocurrent is opposite at the upper and lower edges, as shown in Fig. 2(a). Further measurement by switching the magnetization direction reveals that the sign of the edge photocurrent is reversed with the reversal of the magnetization direction, which are consistent with the APNC response [10,34,35] and can be well explained by the mechanism illustrated in Fig. 1(c): first, since the direction of the temperature gradient is opposite when illuminating the upper and lower edges, it will generate a photocurrent with opposite directions at the upper and lower edges; second, when the magnetization direction is reversed, the sign of ANE is also reversed accordingly, resulting in a sign reversal of APNC. These are more clearly illustrated in Fig. S2 of the Supplemental Material [33]. Figure 2(b) shows the SPCM images taken at different temperatures. The SPCM images are measured during the warming process after FC to 25 K. As the temperature increases, the edge photocurrent decreases gradually and is almost indistinguishable from the SPCM image above 130 K, which is quite below the Curie temperature of 177 K. The fact that the edge photocurrent response vanishes at temperatures quite below the Curie temperature is consistent with previous transport measurements of the anomalous Nernst coefficient $S_{xy}$ [21]. Figure 2(c) shows the scanning along the dashed line (as marked in Fig. 2(b)) at different temperatures, and the peak of the edge photocurrent is extracted to obtain the quantitative temperature dependence of the edge photocurrent response (Fig. 2(d)). According to the Shockley–Ramo theorem [10,15], the temperature-dependent response can be understood quantitatively as:

$$I_{ph} = A \iint \boldsymbol{j}_{th} \cdot \nabla\psi dxdy, \quad (1)$$

where $A$ is the perfector that depends on the electrical configuration, $\psi$ is an auxiliary weighting potential determined by solving Laplace's equation within a specific device geometry, and $\boldsymbol{j}_{th}$ is the local thermoelectric current generated by the laser spot, which is described by the anomalous Nernst effect as:

$$\boldsymbol{j}_{th} = -\alpha\nabla T. \quad (2)$$

Here, $\alpha$ is the 2D thermoelectric tensor and $\nabla T$ is the local temperature gradient induced by the laser spot. Therefore, the temperature dependence of $I_{ph}$ originates from the temperature dependence of the physical quantities in Equations (1) and (2). To illustrate this, we also simulate the temperature dependence of $I_{ph}$ by using the temperature dependences of the transport coefficient such as $\alpha$ obtained in the literature [21] and the temperature dependences of $A$ and the temperature elevation due to laser heating calculated from our experimental parameters (details are presented in Supplemental Material Section III [33]). The simulation results are also shown in Fig. 2(d), which are consistent with our experimental results and further support the anomalous Nernst effect origin of the edge photocurrent. Besides, we also note that the anomalous photo-Nernst current can be used to image the magnetic domain walls at the interface between two magnetic domains in $Co_3Sn_2S_2$ [36]. The clear image of magnetic domain walls provides further evidence that supports the observed photocurrent response is due to anomalous photo-Nernst effect. Furthermore, we also perform measurements under different wavelength excitations (the results are presented in Supplemental Material Section IV [33]), and the APNC shows a broadband response that is consistent with the thermoelectric response.

To further verify that the edge photocurrent observed in our experiments originates from the anomalous Nernst effect, we perform measurements under an external magnetic field along the *c*-axis of $Co_3Sn_2S_2$ with 532-nm excitation. The SPCM image (Fig. 3(a)) after FC to 25 K with $+M$ magnetization shows a clear edge photocurrent response, which is consistent with the results shown in Fig. 2(a). We then investigate the magnetic field dependence of the edge photocurrent at the position denoted by the blue circle in Fig. 3(a), and the result is shown in Fig. 3(b). The edge photocurrent clearly exhibits hysteresis behavior as a function of the external magnetic field, with a corresponding coercive field of approximately 0.7 T. Besides the hysteresis behavior, the edge photocurrent linearly depends on the external magnetic field within our measurement range. Notably, a significant nonzero photocurrent response is observed in the absence of an external magnetic field, corresponding to zero-field anomalous photo-Nernst effect. This low temperature behavior is distinct from that observed in Dirac semimetals $HfTe_5$ [34] and $ZrTe_5$ [35], which exhibits a zero photocurrent response at zero field, owing to the lack of intrinsic magnetism. The SPCM images measured under a higher magnetic field of $\mu_0 H = 5$ T is shown in Fig. 3(c). Compared with the result measured under 0 T (Fig. 3(a)), the external magnetic field can significantly enhance the edge photocurrent, consistent with the results shown in Fig. 3(b). Furthermore, we also performed experiments at room temperature as comparison .The interesting experimental feature at room temperature, as shown in Fig. 3(d),is that no clear edge photocurrent is observed under external magnetic field up to $\mu_0 H = 5$ T (more data presented in Supplemental Material Section V [33]). This behavior stands in sharp contrast to the results observed at 25 K and in similar studies in the literature [10,34,35], where a clear edge photocurrent was observed by applying an external magnetic field. This stark difference suggests that the formation of topological contribution from the Weyl phase plays a crucial role in the observation of large APNC, imply the underlying topological contribution may dominate the APNC response compared with the magnetic order.

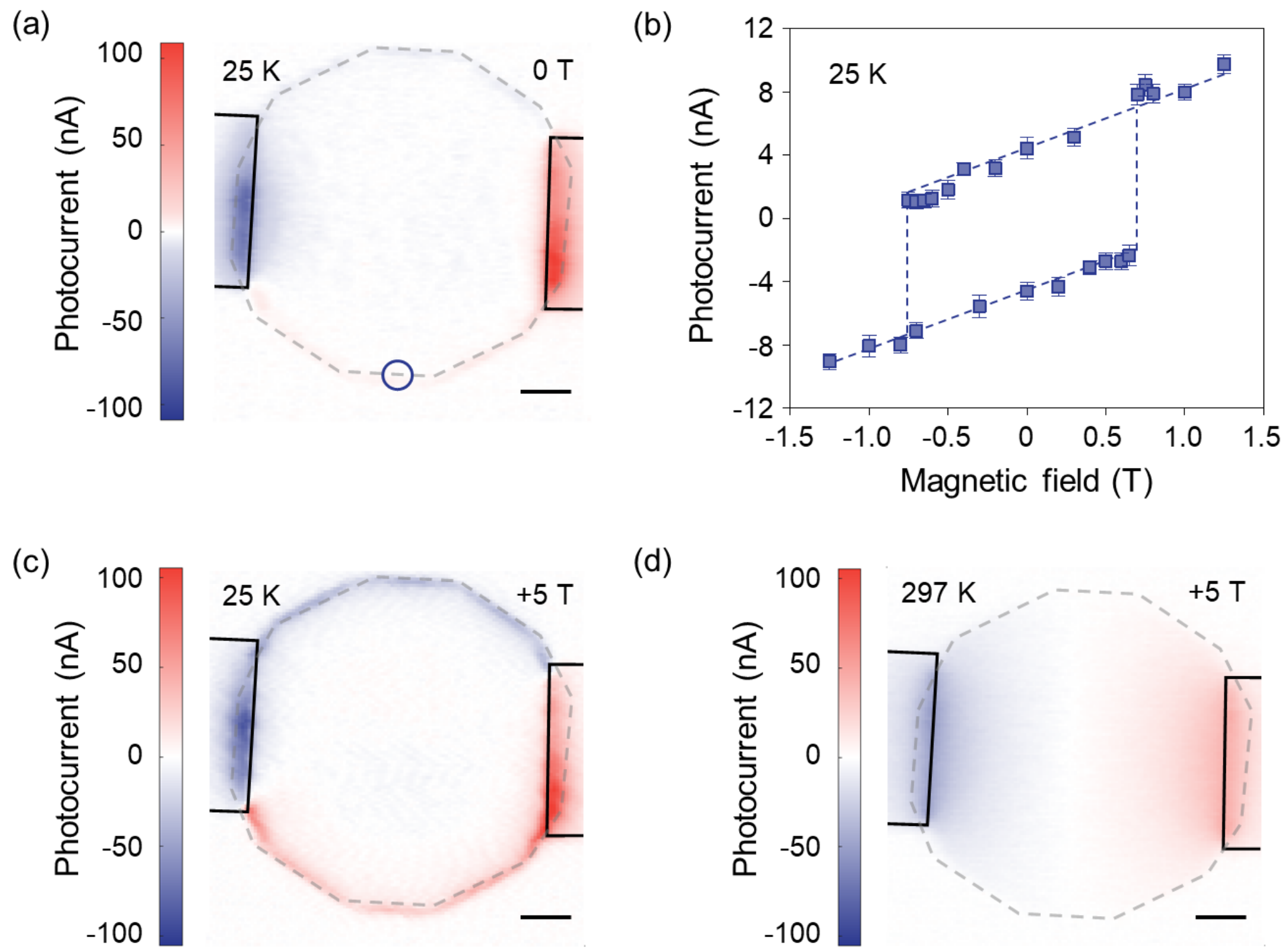

FIG. 3. Anomalous photo-Nernst current measured under an external magnetic field. (a) SPCM images taken at 25 K under zero external magnetic field. The measurement is performed after field-cooling to $+M$. The blue circle denotes the position where the magnetic field dependence measurement in (b) is performed. (b) Magnetic field dependence of the anomalous photo-Nernst current at 25 K. The dashed line is a guide for the eye. The error bars represent the standard deviation obtained from 20 consecutive scans. (c), (d) SPCM images taken under an external magnetic field of $\mu_0 H = 5$ T at 25 K (c) and 297 K (d), respectively. All scale bars are 10 μm. All the measurements are performed under 532-nm excitation with an excitation power of 520 μW.

### B. Enhanced APNC response in the MIR region

In the next, we investigate the wavelength dependence of the APNC by comparing the nominal anomalous Nernst coefficient under excitations at different wavelengths. According to Equations (1) and (2), the relative anomalous Nernst coefficient under excitations at different wavelengths can be calculated using the measured APNC and the simulated temperature profile due to laser heating. In principle, since the photocurrent measurement is a steady-state measurement, the device will reach a steady state with an elevated carrier temperature when the electron temperature reaches equilibrium with the lattice temperature. The steady-state carrier temperature is independent of the excitation wavelength as long as the absorbed light power is the same. Figure 4(a) shows the average temperature elevation within the laser spot under different excitation wavelengths (calculation details are presented in Supplemental Material Section VI [33]), and the temperature variation is less than 8 K over the measurement range. The corresponding upper bounds of the anomalous Nernst coefficients obtained through transport measurements of anomalous Nernst coefficients [21] are also plotted in Fig. 4(a). According to Fig. 4(a), if we consider only the contribution from the final steady state, we should observe a slightly larger value of $S_{xy}$ with 532-nm excitation than with 8-μm excitation. However, as shown in Fig. 4(b), the calculated nominal anomalous Nernst coefficient (calculation details are presented in Supplemental Material Section VI [33]) varies dramatically with the excitation wavelength and significantly increases in the MIR region. This dramatic variance is different from the slight variance predicted in Fig. 4(a), suggesting that the steady-state approximation does not fully capture the anomalous photo-Nernst contribution in this work.

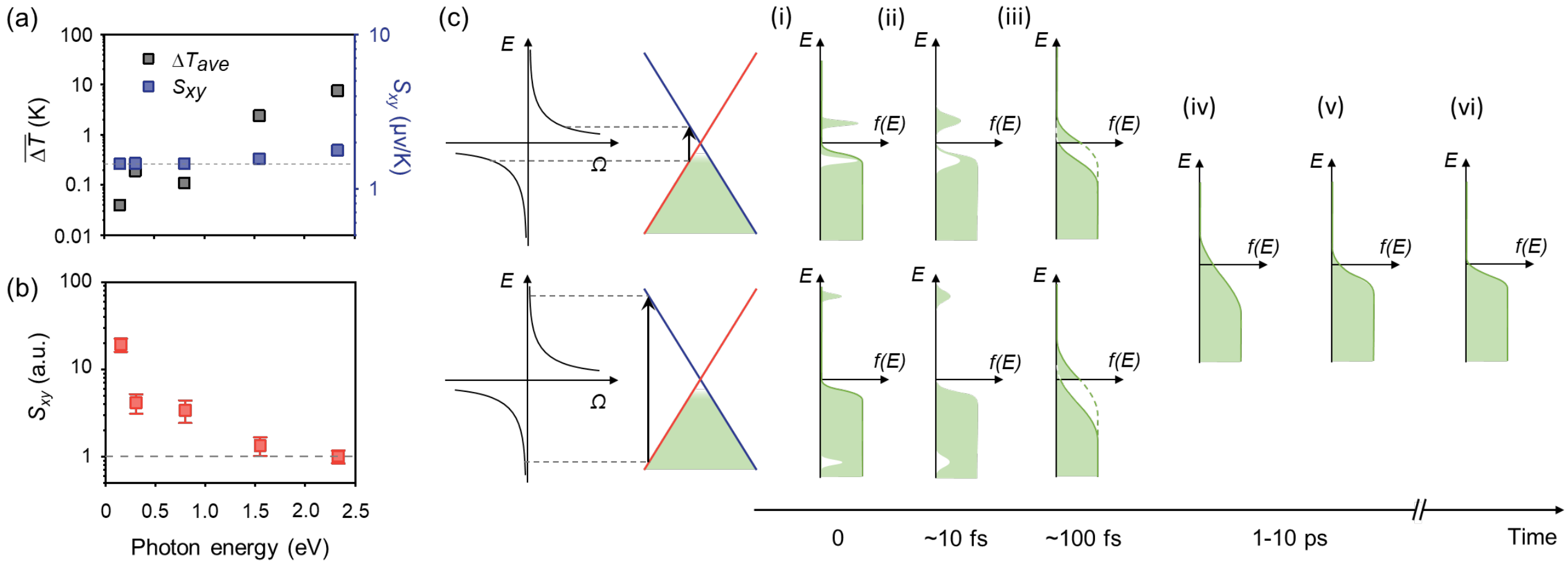

FIG. 4. Enhancement of the APNC response in the MIR region. (a) The average elevated temperature within the laser spot under different excitation photon energies (black squares) and the upper bounds of the anomalous Nernst coefficient (obtained from transport measurements) at different excitation photon energies due to different elevated temperatures (blue squares). The dashed line marks the anomalous Nernst coefficient at 25 K. (b) The nominal anomalous Nernst coefficient under different excitation photon energies, which is calculated according to Equation (1). (c) Schematic illustration of the transient carrier relaxation processes under low-photon-energy (upper panel) and high-photon-energy (lower panel) excitation through carrier–carrier and carrier–phonon scattering before reaching the steady state. The distribution of the Berry curvature in the vicinity of the Weyl cone is shown in the left panel. (i) Initial distribution of carriers after optical excitation, with the green solid line representing the Fermi–Dirac distribution without photoexcitation. (ii) Carrier distribution after initial thermalization through carrier–carrier scattering. (iii) Carriers form two separate uniform distributions within each band. (iv) Carrier distribution after thermalization through interband scattering processes, resulting in a single Fermi–Dirac distribution. (v) Carrier distribution after optical phonon scattering, with a carrier temperature higher than the lattice temperature. (vi) Carrier distribution after achieving equilibrium between the carriers and the lattice through acoustic phonon scattering.

To interpret the dramatic increase in the nominal anomalous Nernst coefficient in the MIR region, we must consider the intermediate unequilibrium state and topological contribution from the Weyl bands to the thermoelectric response. Figure 4(c) illustrates the typical carrier relaxation process after optical excitation [37-41]. Here, we note that whether thermal equilibrium is reached between electrons and holes to form a unified carrier distribution or whether the electrons and holes reach their individual thermal equilibrium first depends on the exact optical phonon scattering time and carrier–carrier scattering time [39,42]. In the plot of Fig. 4(c), we show the scenarios where the unified carrier distributions are reached first. However, it does not affect our analysis either way. According to Fig. 4(c), the intermediate unequilibrium states are very different for low/high excitation photon energies (processes i-iii), although they reach the same steady states (processes iv-vi). In the next, we show how such differences in the intermediate unequilibrium states affect the thermoelectric response through different topological contributions of Weyl bands.

The topological contribution to thermoelectric transport was first described by Xiao et al. in 2006 [5], they pointed out that there is a Berry-phase correction to the magnetization that enters the transport current produced by the thermoelectric response. According to their theory, the in-plane anomalous Nernst conductivity $\alpha_{xy}$ can be expressed as:

$$\alpha_{xy} = \frac{e}{T\hbar}\sum_n \int \frac{d\boldsymbol{k}}{(2\pi)^3}\Omega_{n,z}(\boldsymbol{k})\{(\varepsilon_{n,k}-\mu)f_{n,k} + k_B T \ln[1+e^{-(\varepsilon_{n,k}-\mu)/k_B T}]\}, \quad (3)$$

where $\Omega_{n,z}(\boldsymbol{k})$ is the $z$-component of the $n$-th band's Berry curvature, $\varepsilon_{n,k}$ is the electron energy, $\mu$ is the chemical potential, and $f_{n,k}$ is the Fermi–Dirac distribution. According to Equation (3), $\alpha_{xy}$ is directly related to the Berry curvature of the occupied bands. Therefore, although the final steady states (processes iv-vi of Fig. 4(c)) are the same, the transient processes (processes i-iii of Fig. 4(c)) vary for different wavelengths, and the Berry curvatures of the occupied bands in the transient state differ dramatically at different excitation wavelengths. Consequently, the very different transient states change the nominal anomalous Nernst conductivity $\alpha_{xy}$ at different excitation wavelengths.

The major differences in transient states at different excitation wavelengths can be categorized into two parts. Both parts are related to the Berry curvature enhancement of the occupied states during the transient process. The major part is related to different energy distributions during the transient process. The low-energy photon excitation can effectively induce transitions near the Weyl points comparing to high-energy photon excitation. Since the Berry curvature diverges near the Weyl points, these lower energy carriers experience a significantly larger Berry curvature than those excited by high-energy photons, as shown in Fig. 4(c). Therefore, during the relaxation process, the change in the anomalous Nernst coefficient is significantly greater under low-energy photon excitation, leading to the topological enhancement of the anomalous Nernst coefficient. The second part is simply due to different photon numbers at different excitation wavelengths. With the same absorbed light power, which dumps the same amount of heat to $Co_3Sn_2S_2$, the number of incident photons is larger under MIR excitation because of the much lower photon energy, which will generate more electron–hole pairs in the intermediate state. According to Equation (3), this will lead to a larger anomalous Nernst coefficient. However, after considering the very different spot sizes and subsequent photon number densities at different wavelengths, the contribution from the second part is insufficient to contribute the observed enhancement in the nominal anomalous Nernst coefficient, suggesting the large enhancement in the mid-infrared is dominated by the contribution from the first part with topological origin (see Supplemental Material Section VI [33]). This is consistent with previous transport measurements of anomalous Nernst coefficients, in which $Co_3Sn_2S_2$ exhibited the largest zero-field anomalous Nernst coefficient because of topological enhancement of the Weyl bands [21,24,31] (a detailed comparison is presented in Supplemental Material Section VII [33]). The topological contribution to the anomalous Nernst coefficient distinguishes $Co_3Sn_2S_2$ from ferromagnetic materials. This experimental evidence of the contribution from Berry curvature to the anomalous Nernst effect suggests the interesting possibility of topological engineering of the photothermoelectric conversion in topological materials.

## III. SUMMARY

Notably, the observation of the zero-field anomalous photo-Nernst effect is largely due to the exceptional properties of the magnetic WSM $Co_3Sn_2S_2$. There are at least three deterministic features of $Co_3Sn_2S_2$ that enable the experimental observation of anomalous photo-Nernst currents in this work. First, the conducting nature of the semimetal allows a long-range photocurrent response according to the Shockley–Ramo theorem [15]. This advantage is usually absent for traditional semiconductors, as very high mobility materials or special device structures are required to collect the global current in a semiconducting material [13]; otherwise, the scattering of the carriers limits the observation of a long-range photocurrent even if it is generated locally by the anomalous photo-Nernst effect. Second, the ferromagnetic feature of $Co_3Sn_2S_2$ allows the observation of the anomalous current with a zero field. Among topological semimetals, the experimental observation of the anomalous photo-Nernst effect in Dirac semimetals typically requires the application of an external magnetic field [34,35]. However, as a ferromagnetic material, the anomalous Nernst effect arises from the breaking of the TRS due to the intrinsic magnetization in $Co_3Sn_2S_2$, enabling the observation of the anomalous photo-Nernst effect in the absence of an external magnetic field. Third, topological semimetals typically exhibit larger anomalous Nernst coefficients, which are enhanced by their topological band structures, than traditional semimetals. Consequently, these features in $Co_3Sn_2S_2$ enable the observation of a long-range photocurrent generated by the anomalous photo-Nernst effect.

On the application side, the anomalous photo-Nernst current, benefiting from its zero-field nature, offers an efficient and convenient approach to convert photothermal energy into electrical energy over a broadband range. In contrast to the longitudinal Seebeck effect, where the contributions from electrons and holes typically cancel each other out, the electrons and holes typically deflect in opposite directions in the transverse Nernst effect, and their contributions to the Nernst effect are added up. This results in a simpler application configuration of the Nernst effect than that of the Seebeck effect, where both n- and p-type materials are required to form a thermopile [21]. Furthermore, since the anomalous Nernst coefficient is typically directly related to the topology of the band structure, it can be strongly enhanced by the large Berry curvature in magnetic WSMs, which can further improve the conversion efficiency. On the other hand, the relationship between the anomalous Nernst coefficient and topological properties also makes the anomalous photo-Nernst current a reliable method for probing these topological properties as well as their evolution under various external fields. In summary, our work demonstrates that the anomalous photo-Nernst current arises from the interplay among light, magnetism, and topological order, which is not only interesting for fundamental physics but also holds unprecedented potential for various quantum device applications.

## Acknowledgments

This project was mainly supported by the National Key Research and Development Program of China (Grant Nos. 2021YFA1400100) and the National Natural Science Foundation of China (Grant Nos. 62325401 and 12034001). The authors would also like to acknowledge the support from by the National Key Research and Development Program of China (Grant Nos: 2020YFA0308800 and 2019YFA0704900), the National Natural Science Foundation of China (Grant Nos. 52088101,62250065, and 62227822), and the Open Fund of State Key Laboratory of Infrared Physics (Grant No. SITP-NLIST-ZD-2023-02).

## Competing interests

The authors declare no competing interests.

# Supplemental Material

# Observation of Berry Curvature-Enhanced Anomalous Photo-Nernst Effect in Magnetic Weyl Semimetal

Zipu Fan,[1,*] Jinying Yang,[2,3,*] Yuchun Chen,[1] Ning Zhao,[4] Xiao Zhuo,[1] Chang Xu,[1] Dehong Yang,[1] Jun Zhou,[4] Jinluo Cheng,[5] Enke Liu,[2] and Dong Sun[1,6,7,†]

[1]International Center for Quantum Materials, School of Physics, Peking University, Beijing 100871, China

[2]Beijing National Laboratory for Condensed Matter Physics, Institute of Physics, Chinese Academy of Sciences, Beijing 100190, China

[3]School of Physical Sciences, University of Chinese Academy of Sciences, Beijing 100190, China

[4]School of Physics and Technology, Nanjing Normal University, Nanjing 210023, China

[5]GPL Photonics Laboratory, Key Laboratory of Luminescence Science and Technology, Chinese Academy of Sciences & State Key Laboratory of Luminescence Science and Applications, Changchun Institute of Optics, Fine Mechanics and Physics, Chinese Academy of Sciences, Changchun 130033, China

[6]Collaborative Innovation Center of Quantum Matter, Beijing 100871, China

[7]Frontiers Science Center for Nano-optoelectronics, School of Physics, Peking University, Beijing 100871, China

[*]These authors contributed equally: Zipu Fan, Jinying Yang

[†]Email: sundong@pku.edu.cn

**Table of Contents:**

## I. Materials and Methods

1.1 Sample growth, device fabrication and basic characterization

The chemical vapor transport (CVT) method is used to grow $Co_3Sn_2S_2$ nanoflakes with different thicknesses from 200 to 10 nm. The polycrystalline material is sealed in a quartz tube under high vacuum and used as a precursor. A high-quality single-crystal sample with a good hexagonal morphology is selected to fabricate the device for measurement. Standard electron beam lithography is employed to fabricate the electrode pattern. Ti/Au is deposited via electron beam deposition for electrical contact.

A typical transport measurement result of our $Co_3Sn_2S_2$ nanoflake is shown in Fig. S1. The Curie temperature ($T_C$) of $Co_3Sn_2S_2$ nanoflake with the thickness of 82 nm is 178 K (Fig. S1(a)), which is consistent with previous reports [1]. The anomalous Hall effect is observed below $T_C$ and vanishes when the temperature is higher than $T_C$ (Fig. S1(b)). The positive magnetoresistance and non-linear Hall resistivity indicate the existence of two types of carriers in this sample (Figs. S1(c) and (d)). The carrier density ($n$) and mobility ($\mu$) of holes (h) and electrons (e) at 10 K are found to be $n_h$=1.7×$10^{20}$ $cm^{-3}$, $n_e$=7.8×$10^{19}$ $cm^{-3}$, $\mu_h$=377 $cm^2V^{-1}s^{-1}$, $\mu_e$=609 $cm^2V^{-1}s^{-1}$, as determined by two-carrier fitting [1].

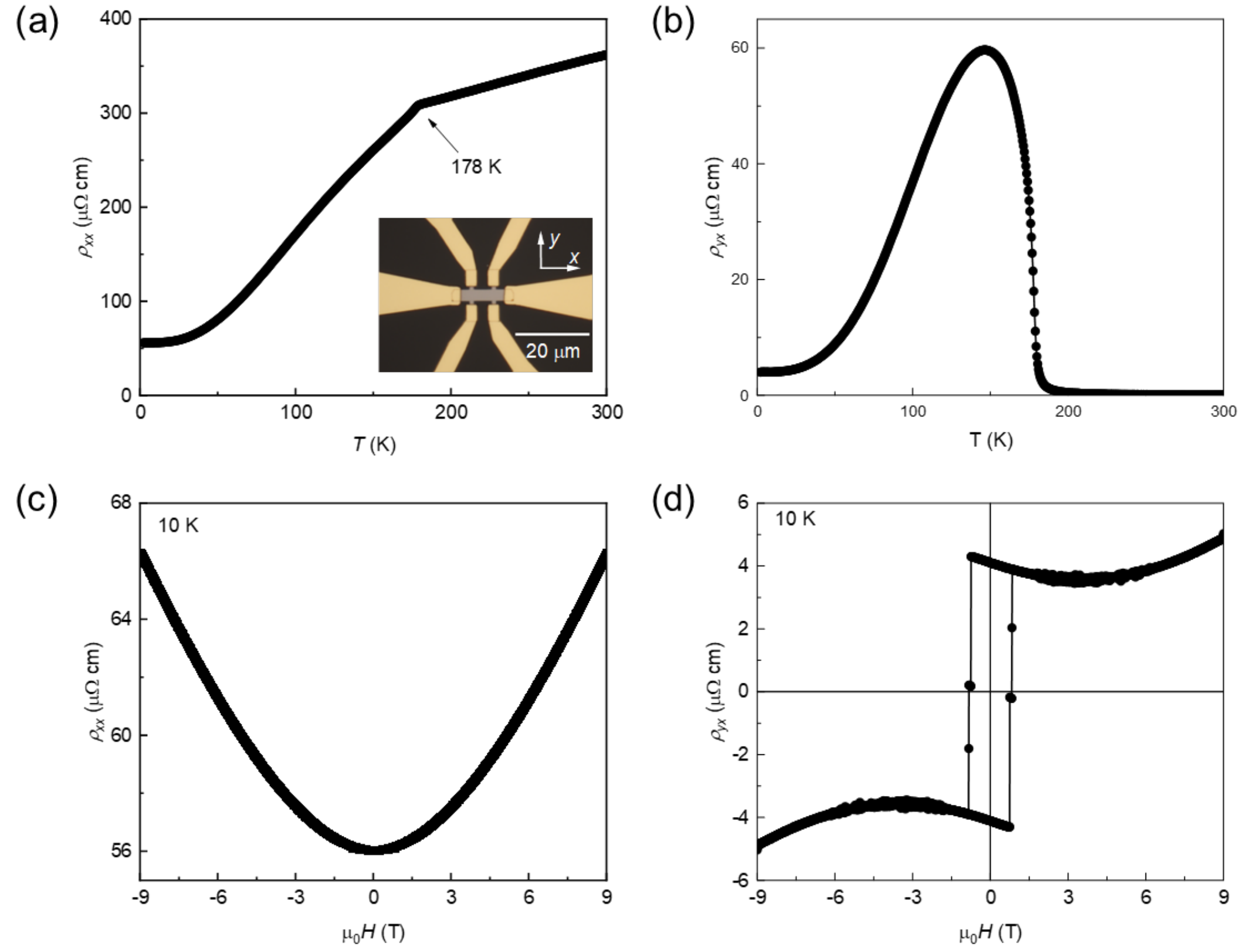

FIG. S1. Transport measurement of the $Co_3Sn_2S_2$ nanoflake. (a), (b) Temperature dependence of resistivity and anomalous Hall resistivity under a magnetic field of 0.1 T. The measurement results have been processed for symmetry and antisymmetry, using the measurement results under magnetic field of ±0.1 T. The current flows along the *x*-axis, and the magnetic field is perpendicular to the sample plane, as shown in the inset of (a). (c), (d) Magnetic field dependence of resistivity and Hall resistivity.

### 1.2 Magnetization process

In the field cooling process, the magnetization of the sample in the experiment is achieved through different cooling cycles in the presence of a magnetic field. The magnetic field is provided by a NdFeB magnet, which is placed on the optical window directly above the device. The magnetic field strength is above 500 Gauss at the location where the device is installed. The magnetization direction is determined by the orientation of the NdFeB magnet.

### 1.3 Photocurrent measurement

For the photocurrent measurements, continuous-wave light from 8-μm, 4-μm, 1550-nm, 800-nm and 532-nm lasers is focused to spot sizes of approximately 16 μm, 8 μm, 14 μm, 2 μm and 1 μm, respectively. The laser beam is modulated via a mechanical chopper (331 Hz), and the photocurrent signal is detected via a current preamplifier (DL Instruments 1211) and a lock-in amplifier (Stanford Research systems SR830). For the scanning photocurrent microscopy measurements, the laser spot is scanned over the device via a galvanometer scanner, and the photocurrent at each position is recorded to form a spatial map. In the magnetic field dependence measurements, a large-bore cryogen-free pulse-tube superconducting magnet is used to apply an external magnetic field of up to 5 T along the *c*-axis of the sample that is placed in a liquid-helium cooled cryostat.

## II. Photocurrent distribution after symmetrization and antisymmetrization

We plot the *M*-symmetric and antisymmetric components of the photocurrent distribution in Fig. S2, which are defined as $I_s = I_{ph}(+M)+I_{ph}(-M)$ and $I_{as} = I_{ph}(+M) - I_{ph}(-M)$. The photocurrent distribution of the symmetric part is very similar to that of the SPCM measured at room temperature, indicating that the photocurrent at the interface with the electrode is not related to magnetization. The antisymmetric component, corresponding to the anomalous photo-Nernst effect, only appears at the edge of the sample, which is consistent with the diagram illustrated in Figs. 1(c) and 1(d) of the maintext.

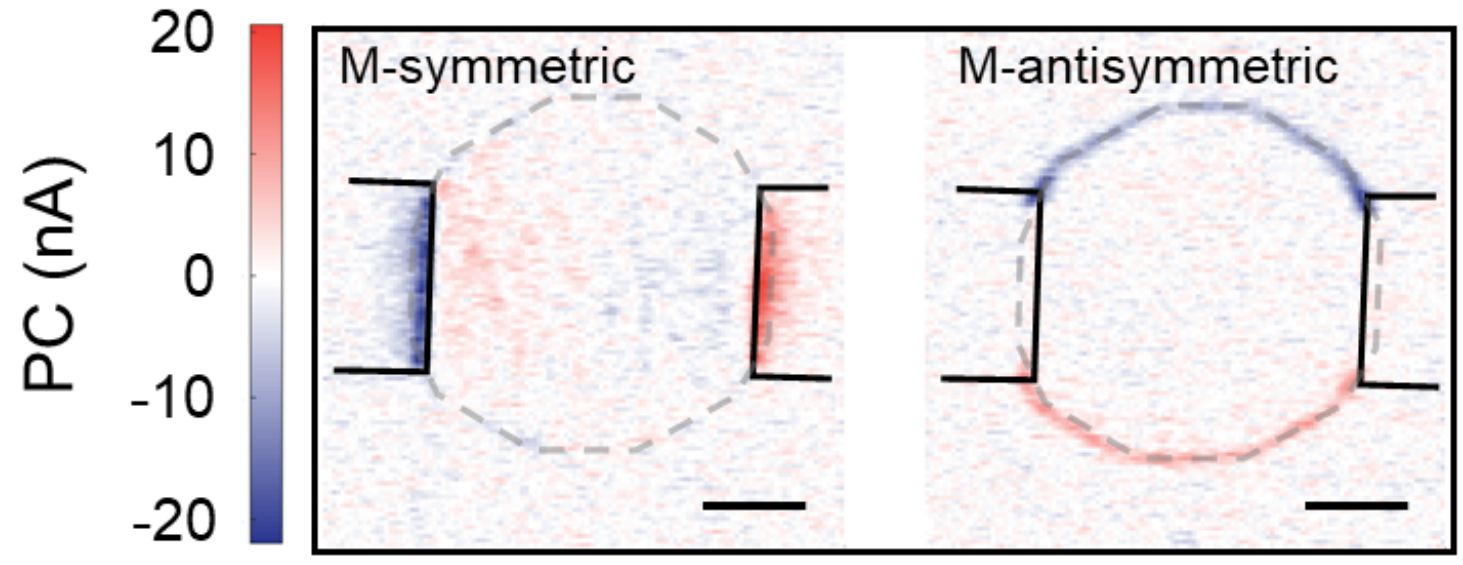

FIG. S2. *M*-symmetric (left panel) and *M*-antisymmetric (right panel) components of the photocurrent data shown in Fig. 2(a) of the maintext. All scale bars are 20 μm.

## III. Simulation of the temperature dependence of APNC

In this section, we present the simulation of the temperature dependence of the anomalous photo Nernst current (APNC). As demonstrated in the main text, the APNC collected by the contacts can be calculated as [2,3]:

$$I_{ph} = A \iint \boldsymbol{j}_{th} \cdot \nabla\psi dxdy, \tag{S1}$$

where $A$ is the prefactor, $\psi$ is the auxiliary weighting potential and $\boldsymbol{j}_{th}$ is the local thermoelectric current:

$$\boldsymbol{j}_{th} = -\alpha\nabla T. \tag{S2}$$

Here, $\alpha = \begin{pmatrix} \alpha_{xx} & \alpha_{xy} \\ \alpha_{yx} & \alpha_{yy} \end{pmatrix}$ is the 2D thermoelectric tensor, and $\nabla T$ is the local temperature gradient induced by the laser spot. Theoretically, the temperature dependence of $I_{ph}$ can be obtained by performing simulations at different temperatures via Equations (S1) and (S2). However, this requires a significant computational effort. In fact, Equations (S1) and (S2) can be further simplified by assuming a specific device structure. As illustrated [2], for a rectangular-shaped sample (Fig. S3), $I_{ph}$ can be expressed as:

$$I_{ph} \sim AS_{xy}\rho_{xx}^{-1}\overline{\Delta T_l}, \tag{S3}$$

where $A$ is the prefactor depending on the electrical measurement circuit, $S_{xy}$ represents the anomalous Nernst coefficient, $\rho_{xx}$ is the longitudinal resistivity, and $\overline{\Delta T_l}$ is the average lateral temperature difference induced by the laser illumination between the upper and lower edges [2]:

$$\overline{\Delta T_l} = \frac{1}{L}\int_0^L dx \int_0^W \frac{\partial T}{\partial y} dy. \tag{S4}$$

where L and W denote the length and width of the rectangular-shaped sample, respectively. Since the simulation results are not expected to strongly depend on the sample shape, we performed the simulation on a rectangular sample for simplicity, as shown in Fig. S3.

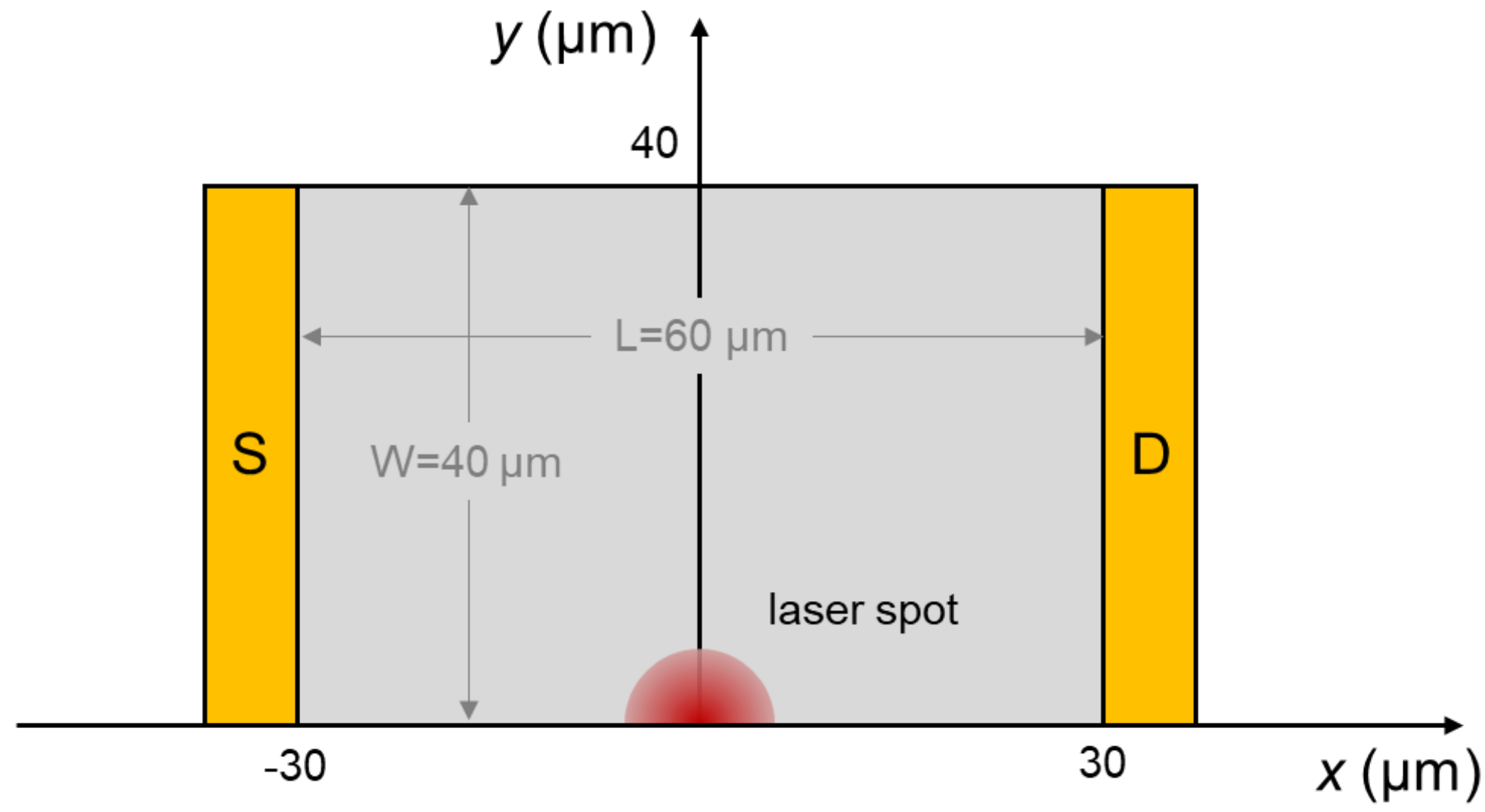

FIG. S3. Schematic illustration of the device geometry used for the simulation. The sample dimensions are 60 μm × 40 μm, and the laser spot is set to the center at position (0,0).

According to Equation (S3), the temperature dependence of $I_{ph}$ originates from the temperature dependences of $S_{xy}$, $\rho_{xx}$, $A$ and $\overline{\Delta T_l}$. In the simulation, the temperature dependence of $S_{xy}$ and $\rho_{xx}$ is acquired from the transport measurement results reported in the literature [4]. Below, we present the experimental and simulation results of the temperature dependences of $A$ and $\overline{\Delta T_l}$, respectively.

### 3.1 Temperature dependence of $A$

According to the Shockley–Ramo theorem [3], the prefactor $A$ can be calculated as

$$A = R_{dev}/(R_{dev} + R_{ext}), \quad \text{(S5)}$$

where $R_{dev}$ and $R_{ext}$ represent the resistance of the device and the external circuit, respectively. The equivalent circuit is illustrated in Fig. S4(a). For the temperature dependence of $A$, we measured the temperature dependences of $R_{dev}$ and $R_{ext}$ to account for the temperature dependence of $A$, as shown in Fig. S4(b). Since we can only measure the total resistance of the circuit, the resistance of the external circuit ($R_{ext}$) is measured by short-circuiting the two terminals of the device, and the resistance of the device ($R_{dev}$) is acquired by subtracting the resistance of the external circuit ($R_{ext}$) from the total resistance.

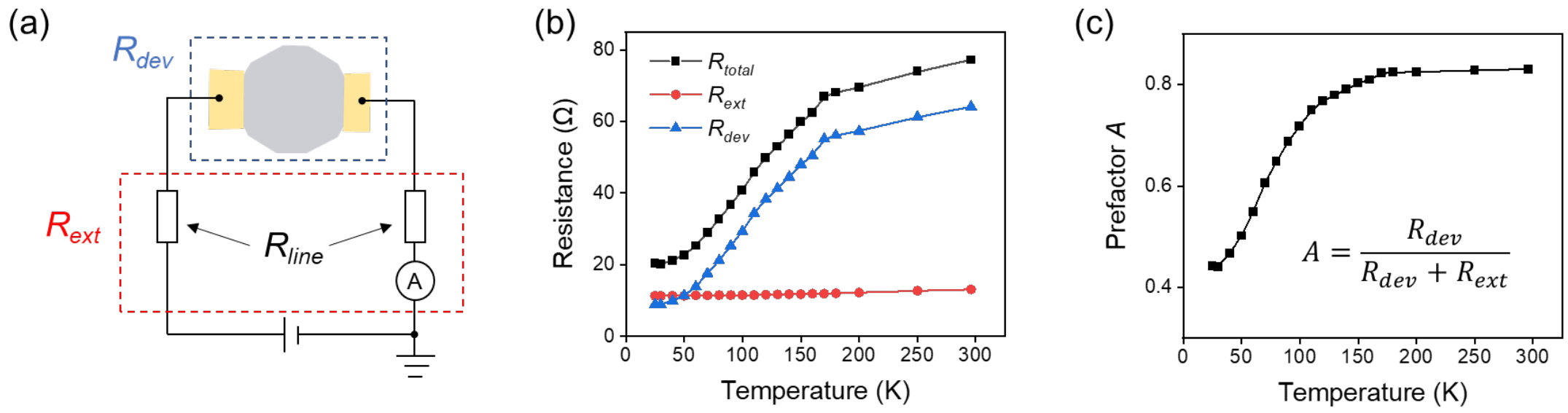

FIG. S4. Temperature dependence of prefactor $A$. (a) Schematic diagram of the equivalent electrical circuit configuration for our current measurement. (b) Temperature dependence of the resistance of the sample and of the external circuit. (c) Temperature dependence of prefactor $A$ calculated from the data in (b).

### 3.2 Temperature dependence of $\overline{\Delta T_l}$

To obtain the temperature dependence of $\overline{\Delta T_l}$, we need to numerically simulate the local temperature profile $T(x, y, z)$ of $Co_3Sn_2S_2$, which involves modeling the heat transfer process within the device. The heat transfer model used to simulate the local temperature increase due to laser heating is shown in Fig. S5. In the model, the heat transfer process includes two steps: (1) The light absorption heats the $Co_3Sn_2S_2$ layer through carrier absorption, and the heat is conducted within the $Co_3Sn_2S_2$ layer both laterally and vertically; (2) Subsequently, the heat transfers across the $Co_3Sn_2S_2/Al_2O_3$ interface into the $Al_2O_3$ substrate. Here, we do not need to consider the heat conduction process within the $Al_2O_3$ substrate. Since the thermal conductivity of $Al_2O_3$ [5] ($10^4$ W K$^{-1}$ m$^{-1}$ at 25 K) is approximately three orders of magnitude greater than that of $Co_3Sn_2S_2$ [6] (8 W K$^{-1}$ m$^{-1}$ at 25 K), we can treat the $Al_2O_3$ substrate as a large cold reservoir and maintain its temperature at the experimental temperature $T_0$.

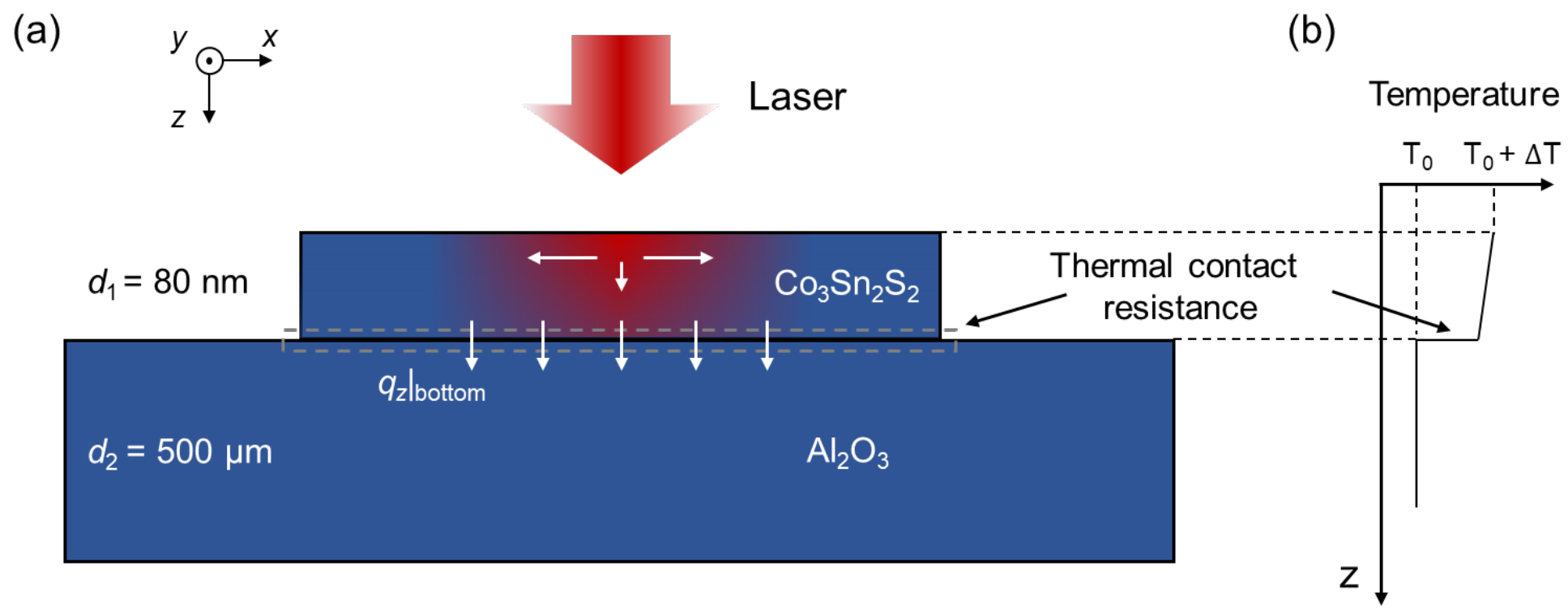

FIG. S5. Schematic of the heat transfer process. (a) Schematic of the heat transfer model used for the simulation of the local temperature profile $T(x,y,z)$. (b) Schematic of the vertical temperature profile.

For step (1), the heat conduction within the $Co_3Sn_2S_2$ layer can be described by the conduction equation [7,8]:

$$\nabla \cdot \boldsymbol{q} = \alpha \mathcal{T} P_0 \frac{1}{2\pi r^2} e^{-\frac{x^2+y^2}{2r^2}} e^{-\alpha z}, \tag{S6}$$

where $\boldsymbol{q} = \boldsymbol{q}(x,y,z)$ is the three-dimensional heat flux, which can be expressed as:

$$\boldsymbol{q} = -\kappa \nabla T. \tag{S7}$$

Here, the right side of Equation (S6) represents the heat source due to light absorption, which is calculated by assuming a Gaussian profile and Beer's law. In the above equations, $\gamma$ is the absorption coefficient, $\mathcal{T}$ is the transmission coefficient at the top surface of $Co_3Sn_2S_2$, $P_0$ is the incident power, $r$ is the radius of the laser spot, $\kappa = \begin{pmatrix} \kappa_{xx} & 0 & 0 \\ 0 & \kappa_{xx} & 0 \\ 0 & 0 & \kappa_{zz} \end{pmatrix}$ is the three-dimensional thermal conductivity tensor of $Co_3Sn_2S_2$, and $T = T(x,y,z)$ represents the local temperature profile.

For step (2), the temperature at the bottom of the $Co_3Sn_2S_2$ layer is determined by the thermal contact resistance through [9]:

$$T(x,y,d_1) - T_0 = R_c q_z(x,y,d_1). \tag{S8}$$

where $T(x,y,d_1)$ represents the temperature at the bottom of the $Co_3Sn_2S_2$ layer, $d_1$ represents the thickness of the $Co_3Sn_2S_2$ layer, $R_c$ represents the thermal contact resistance at the $Co_3Sn_2S_2/Al_2O_3$ interface, and $q_z(x,y,d_1)$ represents the heat transfer from the bottom surface of $Co_3Sn_2S_2$ into the $Al_2O_3$ substrate.

By combining Equations (S6), (S7) and (S8), we can simulate the temperature increase due to laser heating. The parameters used in the simulation are summarized in Table S1. Here, the temperature dependence of the in-plane thermal conductivity ($\kappa_{xx}, \kappa_{yy}$) is taken from the literature[6], while the out-of-plane thermal conductivity is set to be smaller than the in-plane thermal conductivity ($\kappa_{zz} = 2$ W $K^{-1}$ $m^{-1}$) to account for anisotropic heat conduction [10,11]. We also note that owing to the absence of experimental data for the thermal contact resistance ($R_c$) between $Co_3Sn_2S_2$ and $Al_2O_3$, we use a typical value of $R_c = 6 \times 10^{-8} \mathrm{m^2K/W}$ at 40 K in our simulation, which falls within the typical range for $R_c$ ($10^{-8} \sim 10^{-7}$ $\mathrm{m^2K/W}$) between $Al_2O_3$ substrates and other metals [12-14]. For the temperature dependence of $R_c$, according to the acoustic mismatch model and diffuse mismatch model, $R_c$ follows a $T^{-3}$ relationship at extremely low temperatures, whereas at higher temperatures, the trend slows down and eventually saturates [9]. Therefore, we adopt a $T^{-3}$ dependence in the low-temperature regime (25-40 K) and use a $T^{-1}$ approximation to fit the slower trend in the higher temperature range (40-180 K), which is also the case at the graphene/$SiO_2$ interface [15].

**Table S1. Parameters used for the simulation of the temperature profile**

| Thickness of $Co_3Sn_2S_2$ $\boldsymbol{d_1}$ | 80 nm | |
|---|---|---|
| Incident power $\boldsymbol{P_0}$ | 0.36 mW | |
| Spot radius $\boldsymbol{r}$ | 1.6 μm | |
| Transmission coefficient $\boldsymbol{\mathcal{T}}$ | 0.5, Ref. [16] | |
| Absorption coefficient $\boldsymbol{\alpha}$ | 1/20 $\mathrm{nm^{-1}}$, Ref. [17] | |
| Thermal conductivity $\boldsymbol{\kappa}$ | $\kappa_{xx}, \kappa_{yy}$ | 8 $\mathrm{W\,K^{-1}\,m^{-1}}$ (30 K), Ref. [6] |
| | $\kappa_{zz}$ | 2 $\mathrm{W\,K^{-1}\,m^{-1}}$, Ref. [10] |
| Thermal contact resistance $\boldsymbol{R_c}$ | 25-40 K | $\frac{40^3}{T^3} \times 6 \times 10^{-8}$ $\mathrm{m^2\,K\,W^{-1}}$, Ref. [9] |
| | 40-180 K | $\frac{40}{T} \times 6 \times 10^{-8} \mathrm{m^2\,K\,W^{-1}}$, Ref. [15] |

The simulated results of the temperature increase at the center of the laser spot ($\Delta T_{peak} = T(0,0,0) - T_0$) are shown in Fig. S6(a). The simulation is performed via COMSOL. Since we do not have exact values for the parameters $\kappa_{zz}$ and $R_c$ reported in the literature previously, we also performed simulations using several different values (adopting the $T^{-1}$ dependence for $R_c$), and the results are shown in Figs. S6(b) and S6(c), respectively. Although the absolute values of $\Delta T_{peak}$ vary with $\kappa_{zz}$ and $R_c$, both simulations exhibit similar dependences on temperature. The exact values of $\Delta T_{peak}$, however, are not critical to our simulation and are included in an overall fitting coefficient.

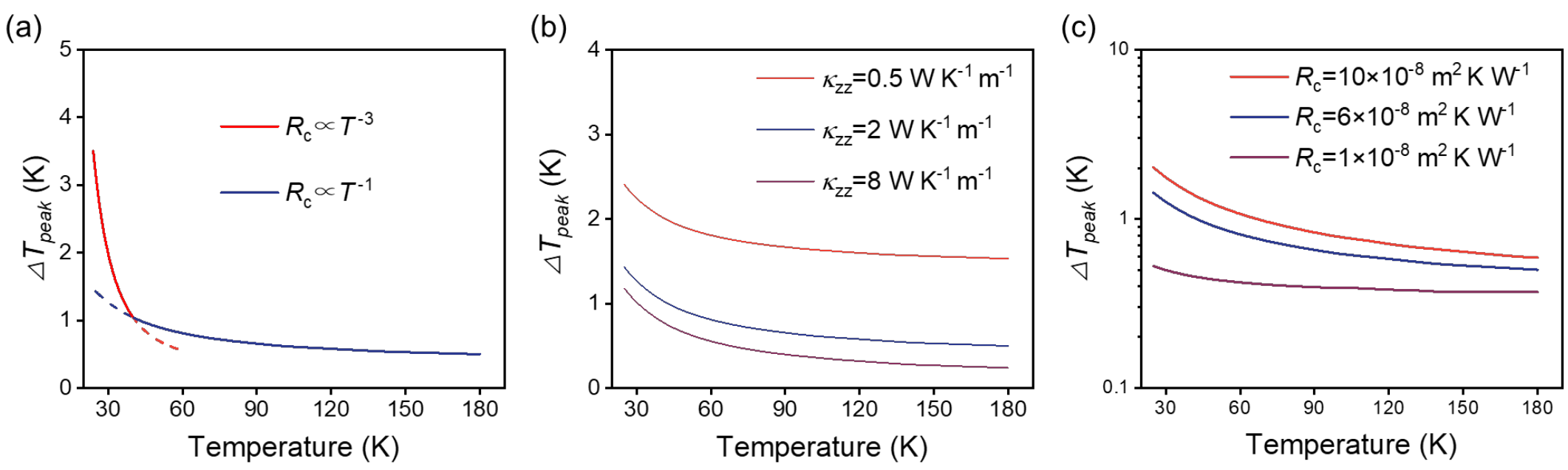

FIG. S6. Simulation results of the temperature dependence of $\Delta T_{peak}$. (a) Simulation result of $\Delta T_{peak}$ using the parameters presented in Table S1. (b) Simulation result of $\Delta T_{peak}$ using different values of $\kappa_{zz}$. The other parameters are presented in Table S1, adopting the $T^{-1}$ dependence for $R_c$. (c) Simulation result of $\Delta T_{peak}$ using different values of $R_c$ at 40 K. The other parameters are presented in Table S1, adopting the $T^{-1}$ dependence for $R_c$.

After the local temperature profile $T(x, y, z)$ is obtained, $\overline{\Delta T_l}$ can then be calculated according to Equation (S2). Finally, on the basis of the temperature dependence of all relevant parameters in Equation (S3), we can simulate the temperature dependence of the anomalous photo Nernst current $I_{ph}$, and the results are shown in Fig. 2(d) of the main text.

## IV. Broadband response of the APNC after the FC process

In this section, we present the SPCM results measured under various wavelength excitations to illustrate the broadband response nature of the anomalous photo-Nernst effect. As shown in Fig. S7, we observed a broadband edge photocurrent response under excitation from 532 nm to 8 μm. The broadband response is consistent with its photothermal origin. Here, we note that the spatial resolutions are different for different excitation wavelengths in Fig. S7 because of the different diffraction limits of the laser spots at different wavelengths, as the excitation wavelengths and the numerical apertures of the objective lenses vary across different measurements. Here, the data measured under 532-nm and 1550-nm excitations are rotated to align with the orientation of the data measured under 4-μm and 8-μm excitations.

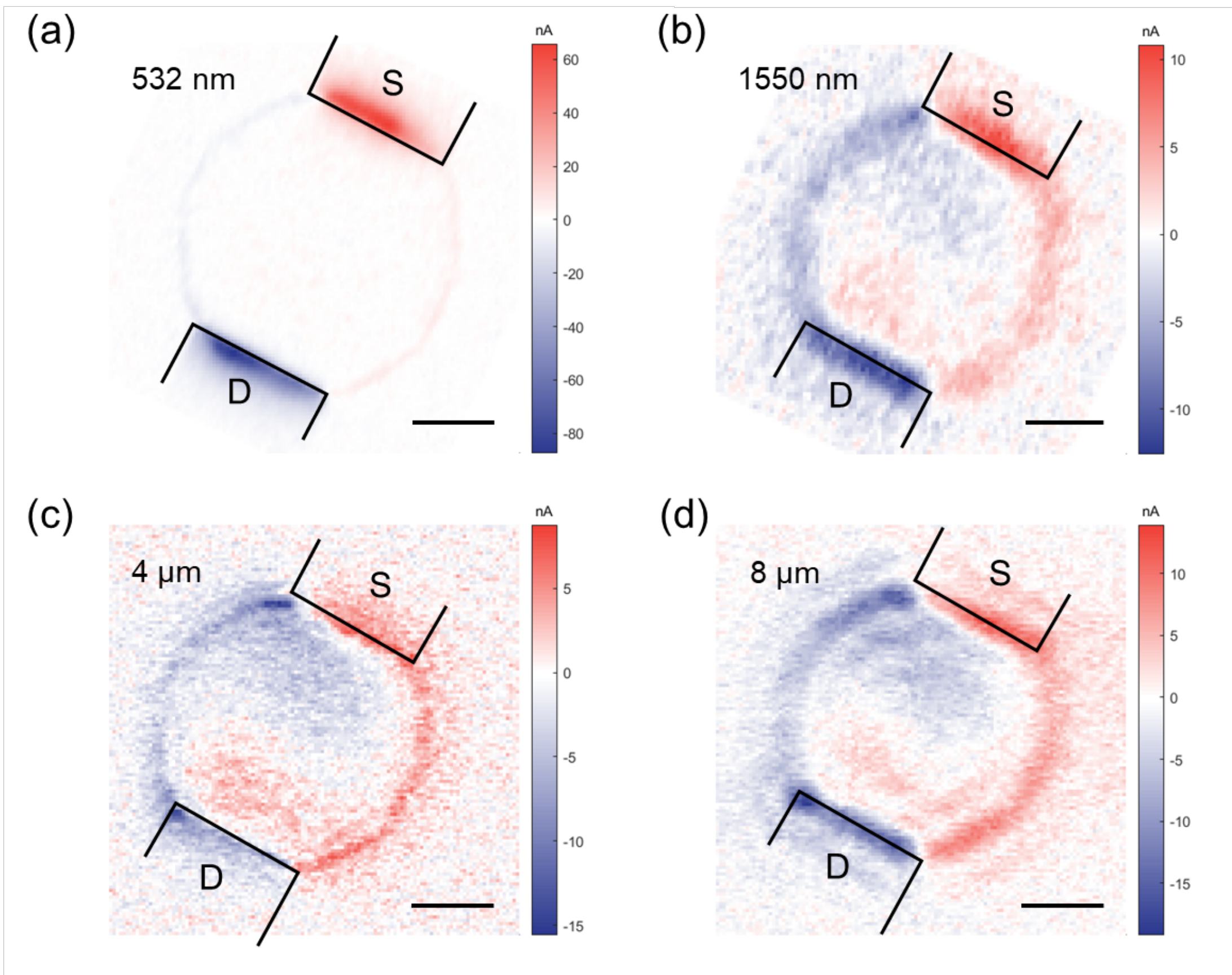

FIG. S7. Broadband response of the APNC. (a)-(d) SPCM images taken at 532 nm (a), 1550 nm (b), 4 μm (c) and 8 μm (d), respectively. The experiments were performed at 25 K after the FC process. All the scale bars are 20 μm.

## V. SPCM images measured at room temperature under an external magnetic field

The scanning photocurrent microscopy (SPCM) images measured at room temperature under an external magnetic field are shown in Fig. S8. However, no clear edge photocurrent is observed, in contrast to the result measured at 25 K (Fig. 2(a) of the main text).

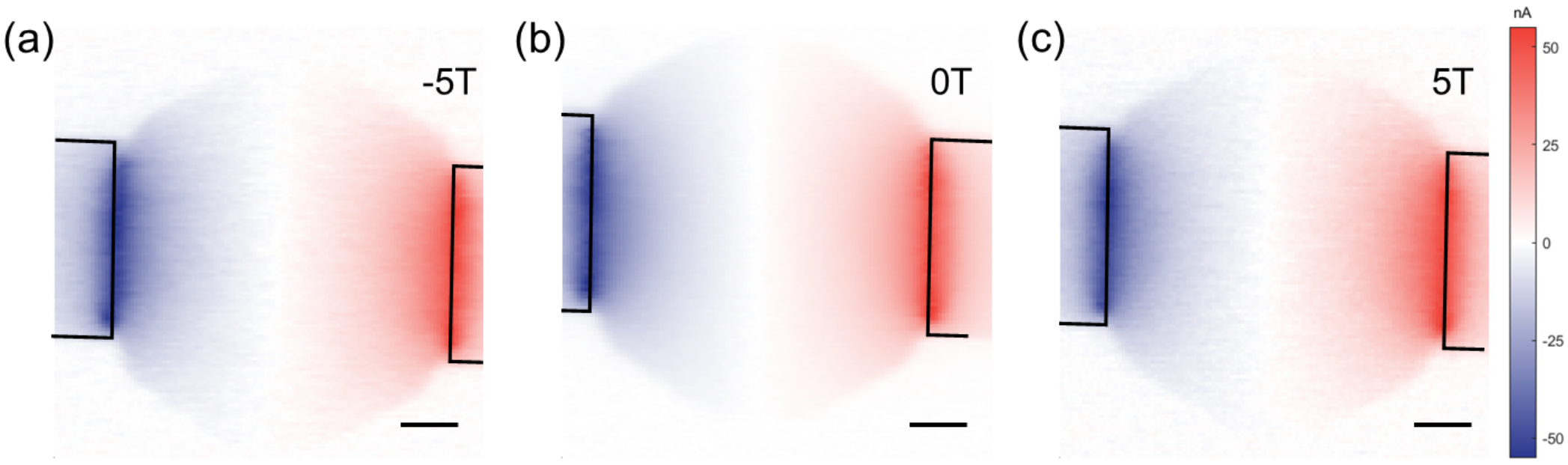

FIG. S8. SPCM images taken at room temperature. (a)-(c) SPCM images taken at 297 K under an external magnetic field of $\mu_0 H$=-5 T (a), 0 T (b), and 5 T (c), respectively. The measurements were performed under 532-nm excitation with an excitation power of 1 mW. All scale bars are 10 μm.

## VI. Comparison of $S_{xy}$ under different wavelength excitations

In this section, we present the calculation details of the nominal anomalous Nernst coefficient $S_{xy}$ under different wavelength excitations. The nominal anomalous Nernst coefficient $S_{xy}$ is calculated according to Equation (S3), $I_{ph} \sim A S_{xy} \rho_{xx}^{-1} \overline{\Delta T_l}$. In this equation, both $A$ and $\rho_{xx}$ do not have any wavelength dependence but should depend on the elevated temperature after light heating at different wavelengths. We show that such variance with temperature is very weak because of the very small temperature elevation range at different excitation wavelengths after we obtain the numerical value of the elevated temperature after light heating. Therefore, only $I_{ph}$, $S_{xy}$ and $\overline{\Delta T_l}$ have wavelength dependences, and the wavelength dependence of $S_{xy}$ can be deduced from the wavelength dependences of $I_{ph}$ and $\overline{\Delta T_l}$. The wavelength dependence of $I_{ph}$ can be obtained directly from the photocurrent measurements shown in Fig. S9 and Fig. 2(a) of the main text. For the calculation of the temperature $T$ under excitations at different wavelengths, we used the same model shown in Section IV. In addition to the parameters listed in Table S1, other wavelength-dependent experimental parameters used for the calculation of $\overline{\Delta T_l}$ are summarized in Table S2.

**Table S2. Parameters used for the calculation of $\overline{\Delta T_l}$**

| Wavelength ($\lambda$) | Excitation power ($P_0$) | Spot radius ($r$) | Absorption coefficient [16,17] ($\alpha$) | Transmission coefficient [16,17] ($\mathcal{T}$) |
|---|---|---|---|---|
| 532 nm | 0.45 mW | 1 μm | 1/16 nm$^{-1}$ | 0.54 |
| 800 nm | 0.36 mW | 1.6 μm | 1/20 nm$^{-1}$ | 0.5 |
| 1550 nm | 1.5 mW | 14 μm | 1/33 nm$^{-1}$ | 0.39 |
| 4 μm | 1.4 mW | 8 μm | 1/83 nm$^{-1}$ | 0.39 |
| 8 μm | 1.65 mW | 16 μm | 1/110 nm$^{-1}$ | 0.29 |

The values calculated $\overline{\Delta T_l}$ under excitation at different wavelengths are shown in Table S3. The wavelength-dependent photocurrent response can be extracted from the experimental data presented in Fig. S9 and Fig. 2(a) of the main text. The nominal anomalous Nernst coefficient $S_{xy}$ can be calculated as the ratio of $I_{ph}$ to $\overline{\Delta T_l}$, with the results presented in Fig. 4(b) of the main text. Here, $\overline{\Delta T_l}$ is used to quantify the temperature gradient that drives the thermal diffusion of carriers. To estimate the variance of parameters such as $S_{xy}$, $A$ and $\rho_{xx}$ due to the temperature elevation, we also calculate the average elevated temperature within the laser spot $\overline{\Delta T}$:

$$\overline{\Delta T} = \frac{1}{\pi r^2 L}\int_0^L dz \iint_{x^2+y^2<r^2} (T - T_0)\,dxdy, \tag{S9}$$

where $r$ is the radius of the laser spot, $L$ is taken as the minimum of the penetration depth ($\alpha^{-1}$) and the sample thickness to account for the light absorption, and $T_0$ is the initial temperature. The calculated results are also shown in Fig. 4(a) of the main text and Table S3.

**Table S3. Simulation results according to Equations (S6) - (S8)**

| $\lambda$ (nm) | $\overline{\Delta T_l}$ (K) | $\Delta A$ | $\Delta\rho_{xx}$ |
|---|---|---|---|
| 532 | 4.3e-1 | 1.5% | 8.0% |
| 800 | 2.1e-1 | 0.1% | 2.2% |
| 1550 | 7.2e-2 | <0.1% | 0.1% |
| 4000 | 8.1e-2 | <0.1% | 0.2% |
| 8000 | 2.3e-2 | <0.1% | <0.1% |

In the above simulation, we assume that the electron temperature equals the lattice temperature. However, there may be a difference between the carrier temperature and the lattice temperature depending on the relative heat coupling strength between the carrier-lattice and heat coupling to the substrate [18]. Next, we estimate the upper bound of the elevated carrier temperature due to this effect. To account for this, the upper bound of the carrier temperature difference between the electron and the lattice would be that all the light power is absorbed by the electrons within the typical electron–optical phonon scattering time of τ, and the heat dissipation to the lattice $Co_3Sn_2S_2$ and the sapphire substrate is ignored during τ. After τ, the elevated temperature decreases by coupling with the lattice and subsequently to the substrate. Typically, the coupling with the substrate is far slower than the carrier–phonon coupling of $Co_3Sn_2S_2$. The carrier–optical phonon scattering time constant is measured to be ~6 ps, and the major optical phonon relaxation lasts less than 25 ps from previously reported ultrafast pump–probe measurements [19]. Under this diagram, the upper bound of the electron temperature can be calculated as:

$$\int_{T_0}^{T_e} \rho C_e(T) dT = P\tau, \quad \text{(S10)}$$

where $\rho = 7.244\ g\ cm^{-3}$ is the density of $Co_3Sn_2S_2$ [1], $C_e(T) = \gamma T$ ($\gamma = 10.8\ mJ\ mol^{-1}\ K^{-2}$) is the heat capacity of electron [20], and $P$ is the absorption power density, which can be expressed as:

$$P = \alpha \mathcal{T} P_0 \frac{1}{2\pi r^2} e^{-\frac{x^2+y^2}{2r^2}} e^{-\alpha z}, \quad \text{(S11)}$$

The parameters of Equation (S11) under different wavelength excitations are shown in Table S2. According to Equations (S10) and (S11), we can simulate the electron temperature under excitation at different wavelengths, and the results are shown in Table S4. Here, the maximum average temperature elevation $\overline{\Delta T}$ is only 9.8 K for 532-nm excitation, which sets the upper bound of the carrier temperature after photoexcitation. Notably, the upper bound of the temperature elevation could account for only a 28% change in the steady-state anomalous Nernst coefficient $S_{xy}$, as shown in Table S4. Therefore, even the upper bound of temperature elevation cannot account for the dramatic variance in the nominal $S_{xy}$ at different wavelengths. In addition, the effect of elevated temperature yields an excitation wavelength-dependent trend that is opposite to the experimentally observed results shown in Fig. 4(b). This suggests that the major contribution of the observed enhancement of the nominal anomalous Nernst coefficient in the mid-infrared region is not due to different elevated temperatures at different excitation wavelengths.

**Table S4. Simulation results of the electron temperature**

| Wavelength ($\lambda$) | Relaxation time (τ) | | | |
|---|---|---|---|---|
| | 1 ps | 10 ps | 25 ps | |
| | $\overline{\Delta T}$ $(K)$ | $\overline{\Delta T}$ $(K)$ | $\overline{\Delta T}$ $(K)$ | $\Delta S_{xy}$ |
| 532 nm | 4.6e-1 | 4.3 | 9.8 | 27.8% |
| 800 nm | 1.1e-1 | 1.1 | 2.6 | 8.4% |
| 1550 nm | 2.8e-03 | 2.8e-2 | 6.9e-2 | 0.2% |
| 4 µm | 3.1e-03 | 3.2e-2 | 7.9e-2 | 0.2% |
| 8 µm | 5.2e-04 | 5.2e-3 | 1.3e-2 | <0.1% |

Finally, we present a comparison of the absorbed photon density under different wavelength excitations to evaluate the contribution to the anomalous Nernst coefficient due to the different photon number. The absorbed photon density can be calculated as the ratio between the absorbed power density [Equation (S11)] and the photon energy:

$$n(x, y, z) = \alpha \mathcal{T} P_0 \frac{1}{2\pi r^2} e^{-\frac{x^2+y^2}{2r^2}} e^{-\alpha z} / \hbar\omega. \quad \text{(S12)}$$

The average absorbed photon density within the laser spot can be calculated as:

$$\bar{n} = \frac{1}{\pi r^2 L} \int_0^L dz \iint_{x^2+y^2<r^2} n dx dy, \quad \text{(S13)}$$

where $r$ is the radius of the laser spot and $L$ is taken as the minimum of the penetration depth ($\alpha^{-1}$)

and the sample thickness to account for the light absorption. Using the parameters shown in Table. S2, we can calculate the average absorbed photon density under different wavelength excitations. However, the average absorbed photon density is much higher under 532-nm excitation ($3.43 \times 10^{27}$ $cm^{-3}s^{-1}$) compared to that under 8-μm excitation ($6.16 \times 10^{25}$ $cm^{-3}s^{-1}$). This result suggests that contribution from the different photon number at different wavelength alone is insufficient to account for the observed enhancement in the nominal anomalous Nernst coefficient.

## VII. Comparison of the zero-field $S_{xy}$ values of various magnetic materials

In this section, we present a comparison of the zero-field $S_{xy}$ values of various magnetic materials. The anomalous Nernst coefficient $S_{xy}$ (obtained by extrapolating the slope of the high-field data) and the true zero-field value among various magnetic materials are summarized in Table S5. Here, we list only the coefficients of some typical metals and Weyl semimetals but not those of semiconductors since it is difficult to observe long-range anomalous photo-Nernst currents in intrinsic semiconductors. As shown in Table S5, $Co_3Sn_2S_2$ not only possesses a greater anomalous Nernst coefficient than topologically trivial metals but also presents the largest zero-field anomalous Nernst coefficient among the recently discovered magnetic WSMs, making it ideal for observing the zero-field anomalous photo-Nernst effect.

**Table S5. Comparison of $S_{xy}$ among different magnetic WSMs**

| Type | Systems | $S_{xy}$ (μV/K) | Zero-field $S_{xy}$ (μV/K) | Refs. |
|---|---|---|---|---|
| WSMs | $Co_3Sn_2S_2$ | ~5 (70 K) | ~5 (70 K) | [21] |
| | $UCo_{0.8}Ru_{0.2}Al$ | ~18 (30 K) | ~5 (30K) | [22] |
| | $Mn_3Sn$ | ~0.6 (200 K) | ~0.6 (200 K) | [23] |
| | $Co_2MnGa$ | ~6 (300 K) | 0 | [24] |
| | $EuCd_2As_2$ | ~5.3 (25 K) | 0 | [25] |
| | CeAlSi | ~0.4 (10 K) | 0 | [26] |
| Metals | Fe | ~0.5 (300 K) | / | [27] |
| | Co | ~0.37 (300 K) | / | [27] |
| | Co/Ni | ~0.19 (300 K) | ~0.19 (300 K) | [28] |
| | Pt/Fe | ~1.0 (300 K) | 0 | [29] |